\documentclass[12pt]{report}

\usepackage[utf8]{inputenc}
\usepackage[english]{babel}
\usepackage[T1]{fontenc} 

\usepackage{graphicx} 
\usepackage{float} 
\usepackage{hyperref} 
\usepackage{a4wide} 
\usepackage[parfill]{parskip}
\usepackage{enumerate}
\usepackage{enumitem}
\usepackage{subcaption}
\usepackage{amsmath} 
\usepackage{amssymb}
\usepackage{amsthm}
\usepackage{mathtools} 
\usepackage{amsfonts}
\usepackage{braket}
\usepackage{mathrsfs}

\usepackage{listings} 

\usepackage[backend=biber,sorting=none]{biblatex}
\usepackage{csquotes} 
\newcommand{\Ham}{\hat{\mathcal{H}}}
\newcommand{\anhi}{\hat{a}}
\newcommand{\crea}{\hat{a}^\dagger}
\newcommand{\exci}{\hat{T}}
\newcommand{\sexci}{\hat{\tau}}

\title{Study of Adaptative Derivative-Assemble Pseudo-Trotter Ansatzes in VQE through qiskit API}
\author{Max Alteg, Baptiste Chevalier, Octave Mestoudjian,Johan-Luca Rossi}
\date{April 2022}

\begin{document}

\maketitle

\tableofcontents

\chapter*{Introduction}

\paragraph{} The solution of the time-independent Schrodinger equation for molecular systems allows for the prediction of chemical properties. Quantum Computers are expected to make possible the simulation of quantum systems more efficiently and accurately than classical ones, including the simulation of molecules. This is due to the exponential size of the Hilbert Space.

The first algorithm which has been proposed to solve the problem of simulating electronic structure on a Quantum Computer is well-known as the Phase Estimation Algorithm (PEA). Despite that PEA gives the exact ground state for a given problem, the long circuit and the complex gates used in this algorithm make it impractical on any current or near-term Quantum Hardware referred as NISQ (Noisy intermediate Scale Quantum) devices. 

In order to answer the problem of NISQ era and allows one to outperform classical computers, Variational Quantum Algorithms (VQAs) were designed. VQAs are hybrid classical-quantum algorithms due to the fact that there are using a quantum hardware for computation while using a classical optimizer for optimization. In this way, VQA algorithms allows the creation of shallower circuits and are thus much more noise resilient. The VQA algorithm we are interested in is the so-called Variational Quantum Eigensolver (VQE) algorithm and was originally designed to simulate electronic structures and to compute the ground state of a given molecule. 

VQE is made of two main components. First, an ansatz which has several tunable parameters and his associated quantum circuit. The ansatz is run on the quantum device and aim to simulate the wavefunction, the parameters of the ansatz will be optimize until the expectation value is minimum meaning that one has found the ground state. The second component is then the classical optimizer, it is run on a classical computer and is the often the same as the one used in Machine Learning. At this point, it is clear that if there is a limiting factor for VQE it can only be the ansatz. 

The very first ansatz that has originally been used is called UCCSD and it is based on Coupled Cluster Theory truncated at single and double excitations. One other way that has been investigated alongside is the Hardware efficient ansatz (HEA) which was designed to be efficiently implemented on NISQ devices but is not as efficient as UCCSD. The main issue considering UCCSD is the large amount of parameters to optimize and this leads us to the introduction of Adaptive Derivative-Assembled Pseudo-Trotter ansatz VQE (ADAPT-VQE) which determines a quasi-optimal ansatz with a minimal number of parameters. 

Unlike ansatzes introduced above, the key point of ADAPT-VQE is to grow the ansatz at every step, by adding operators chosen from a pre-determined pool of operators one-at-a-time, assuring that the maximal amount of correlation energy is recovered at each step. There exists different kind of ADAPT-VQE depending on the starting pool of operators as the fermionic-ADAPT, the qubit-ADAPT or even the qubit excitation based (QEB).

The goal of this project is to implement the different types of ADAPT-VQE mentioned before. After being sure to understand the theoretical background under all of these concepts, we will implement each algorithm using quiskit. We will also compare all of these algorithms on different criterions such as the number of parameters, the accuracy or the number of CNOT gate used on $H_2$ and $LiH$ molecules. Then we will have a small discussion about the results we obtained.

\chapter{Theoretical Background}

\section{Quantum Chemistry}

\subsection{Hamiltonian and Born-Oppenheimer approximation}

Considering a molecule formed by N atoms of atomic number $Z_k$, located at the position $R_k$, the \textit{ab initio molecular Hamiltonian} can be written as \cite{vqerev}:

\begin{equation}
    \label{eq:molh}
    \Ham = \hat{T}_{n} + \hat{T}_e + \hat{V}_{n} + \hat{V}_{ne} + \hat{V}_{ee}
\end{equation}

where $\hat{T}_{n}$ and $\hat{T}_e$ are respectively the kinetic energy of the nuclei and of the electrons, and $\hat{V}_{n}$,$\hat{V}_{ne}$ and $\hat{V}_{ee}$ are the electric potential in between the nuclei together, the nuclei and the electrons and the electrons together.

Regarding to the Born-Oppenheimer approximation, one can neglect the kinematic energy of the nuclei as well as his electric potential since it is much heavier than the electrons and his motion can then be neglected.
Hence one get from equation \ref{eq:molh} the Hamiltonian :
\begin{equation}
    \Ham = \hat{T}_e + \hat{V}_{ne} + \hat{V}_{ee}
\end{equation}
where
\begin{align}
    \hat{T}_e &= - \sum_i \frac{\hbar^2}{2m_e} \nabla_i^2\\
    \hat{V}_{ne} &= - \sum_{i,k} \frac{e^2}{4\pi\epsilon_0}\frac{Z_k}{|r_i-R_k|}\\
    \hat{V}_{ee} &= \frac12 \sum_{i \neq j} \frac{e^2}{4\pi\epsilon_0}\frac{1}{|r_i-r_j|}
\end{align}
with $r_i$ the position of the electron $i$, $m_e$ the mass of the electron, $e$ the elementary charge, $\hbar$ the reduced Plank constant and $\nabla_i$ the Laplace operator for electron $i$. 

\subsection{Spin-orbital and Slater determinant}

Before continuing to develop the hamiltonian, we need a basis to represent the wavefunction into. Basis elements are called \textit{spin-orbitals} and are noted $\phi_j(\xi_j)$, majority of VQE algorithms use the so-called Slater Type Orbitals of minimized size, for instance the minimal STO-3G basis. 

As electrons are fermions with half integer spin, the wavefunction have to be \textit{antisymmetric} meaning that $\psi(\xi_1,\ldots,\xi_k,\ldots,\xi_p,\ldots,\xi_N) = -\psi(\xi_1,\ldots,\xi_p,\ldots,\xi_k,\ldots,\xi_N)$, hence two electron can not be simultaneously in the same quantum state, this is what is called the \textit{Pauli exclusion principle}. A relevant idea is then to represent the function as a determinant or a linear combination of determinants (because determinants respect the above properties). \cite{slater}

For a wavefunction of $n$ occupied orbitals, one can write the \textit{Slater determinant} representing the many-body basis function :
\begin{equation}
    \Psi(\xi_1,\xi_2,\ldots,\xi_n) = \frac{1}{\sqrt{n!}}
\begin{vmatrix}
\phi_1(\xi_1) & \phi_2(\xi_1) & \cdots & \phi_n(\xi_1) \\ 
\phi_1(\xi_2) & \phi_2(\xi_2) & \cdots & \phi_n(\xi_2) \\ 
\vdots & \vdots & \ddots & \vdots \\ 
\phi_1(\xi_n) & \phi_2(\xi_n) & \cdots & \phi_n(\xi_n)
\end{vmatrix}
\end{equation}
with the variables $\xi_j = (r_j,\sigma_j)$ representing both position and spin of the electrons.
For practical reasons, we will denote the basis function has :
\begin{equation}
    \ket{\Psi} = \ket{\phi_1\phi_2\ldots\phi_n}
\end{equation}
where $\phi_j=1$ means the j-th basis function $\phi_j(\xi_j)$ is occupied while $\phi_j=0$ means that it is unoccupied.

The Slater determinant can be used as an \textbf{approximation of the wavefunction} as it is the case in the \textit{Hartree-Fock method} \ref{Hartree-Fock}.
Note that $\braket{\Psi_a|\Psi_b} = \delta_{a,b}$
To be exact, one can also expand the wavefunction in the Hilbert space span by those basis functions as a \textbf{linear combination of those determinants}.

\subsection{First and Second Quantization of the Hamiltonian} \label{First and Second Quantization of the Hamiltonian}

\subsubsection{First Quantization}
Defining the wavefunction in terms of Slater determinants ensure that the antisymmetry property if verified as for any permutation $\sigma$ : $\ket{\sigma(\phi_1\phi_2\ldots\phi_n)} = (-1)^{\pi(\sigma)}\ket{\phi_1\phi_2\ldots\phi_n}$
As we know that this property is already respected, one can first quantize the hamiltonian by projecting into the space spanned by the $\{\phi_i(\xi_i)\}$ and obtain the \textit{one and two body integrals}.
\begin{equation}
    h_{pq} = \braket{\phi_p | \hat{T}_e + \hat{V}_{ne} | \phi_q} = \int d\xi \phi_p^*(\xi) \Big(-\frac{\hbar^2}{2m_e}\nabla^2 - \sum_k \frac{e^2}{4\pi\epsilon_0}\frac{Z_k}{|r-R_k|}\Big)\phi_q(\xi)
\end{equation}
\begin{equation}
    h_{pqrs} = \braket{\phi_p\phi_q | \hat{V}_{ee}| \phi_r\phi_s} = \frac{e^2}{4\pi\epsilon_0}\int d\xi_1d\xi_2 \frac{\phi_p^*(\xi_1)\phi_q^*(\xi_2)\phi_r(\xi_1)\phi_s(\xi_2)}{|r_1 - r_2|}
\end{equation}
And thus the first quantized hamiltonian :
\begin{equation}
    \Ham = \sum_{i=1}^m \sum_{p,q=1}^n h_{pq}\ket{\phi_p^{(i)}}\bra{\phi_q^{(i)}} + \frac12 \sum_{i\neq j}^m \sum_{p,q,r,s=1}^n h_{pqrs}\ket{\phi_p^{(i)}\phi_q^{(j)}}\bra{\phi_r^{(i)}\phi_s^{(j)}}
\end{equation}

The first quantization is not often used, even though the mapping is quite straightforward, it is not adapted to NISQ computers.

\subsubsection{Second Quantization}

For second quantization, we need the operators to ensure the antisymetry property. The operators should allow moving from one basis to another (which was obvious in the first quantization), these are the so-called \textit{fermionic creation} $\crea$ and \textit{annhilation} $\anhi$ operators. Those operators are subject to the algebra defined by following anti-commutation rules:
\begin{align}
    \{\anhi_p,\crea_q\} &= \delta_{p,q}\\
    \{\crea_p,\crea_q\} &= \{\anhi_p,\anhi_q\} = 0
\end{align}

Hence the Slater determinant can be re-written as :
\begin{equation}
    \ket{\Psi} = \prod_i (\crea_i)^{\phi_i} \ket{vac} = (\crea_1)^{\phi_1}(\crea_2)^{\phi_2}\ldots(\crea_n)^{\phi_n} \ket{vac}
\end{equation}
and the \textit{Second quantize hamiltonian} as 
\begin{equation}
    \Ham = \sum_{pq} h_{pq}\crea_p\anhi_q + \frac12 \sum_{pqrs} h_{pqrs}\crea_p\crea_q\anhi_r\anhi_s
\end{equation}

\subsection{Mapping} \label{Mapping}
    
There are several ways to map the fermionic operators to spin-operators and so to qubits which are spin-1/2 objects. Moreover, in general the only operators that can be measured on a Quantum Computer are spin-operators or Pauli operators X,Y,Z.

A map $\mathcal{T} : \mathcal{F}_n \xrightarrow[]{} (\mathbb{C}^2)^{\otimes N}$ from the $n$ orbitals Fock states space $\mathcal{F}_n$ to the operators acting on spin states of $N$ qubits  space $(\mathbb{C}^2)^{\otimes N}$ should preserve the antisymetric behaviour of fermionic operators. Jordan and Wigner showed that such an isomorphism maintaining the algebraic structure does exist. The mapping concerns the hamiltonian operator used for computing the expectation value in the VQE as well as the ansatzes initially defined in fermionic terms (for instance the UCC).

The most used mapping is the \textit{Jordan-Wigner mapping}, in this mapping one qubit corresponds to one n-spin-orbital. The value $\ket{0}$ for a qubit corresponds to the orbital being unoccupied while the value $\ket{1}$ corresponds to the orbital being occupied.

One wants to map the fermionic operators to pauli matrices. Intuitively, the mapping according to the operators should be the following :
\begin{align}
    \crea_j &\xrightarrow[]{} \ket{1}\bra{0}_j = \begin{bmatrix}
        0 & 0\\
        1 & 0\end{bmatrix} = \frac{X_j -iY_j}{2}\\
    \anhi_j &\xrightarrow[]{} \ket{0}\bra{1}_j = \begin{bmatrix}
        0 & 1\\
        0 & 0\end{bmatrix} = \frac{X_j +iY_j}{2}
    \label{eq:JWop}    
\end{align}
where $X_j$ and $Y_j$ act on the j-th qubit. But the antisymetry relation is not respected anymore. To do so, we need to add a Z operators string of length $j$ such that the eigenvalue will be $+1$ for even number of occupied orbitals and $-1$ for odd number of occupied orbitals. The final mapping is then :
\begin{align} \label{eq:jordan-wigner}
    \crea_j &\xrightarrow[]{} \frac{X_j -iY_j}{2} \bigotimes_{k=0}^{j-1} Z_k\\
    \anhi_j &\xrightarrow[]{} \frac{X_j +iY_j}{2} \bigotimes_{k=0}^{j-1} Z_k
\end{align}

The number of pauli string generated in this mapping scale in $\mathcal{O}(N^4)$ where $N$ is the number of qubits.

\section{Ansatzes}
\label{Ansatzes}

\subsection{Introduction : Variational Method}

The main idea of VQE starts from the variational method which is used in physics to solve several problems. For instance, one can use this method to compute \textit{approximation} of ground state and ground energy of molecules.

\begin{figure}[H]
    \centering
    \includegraphics[scale=0.5]{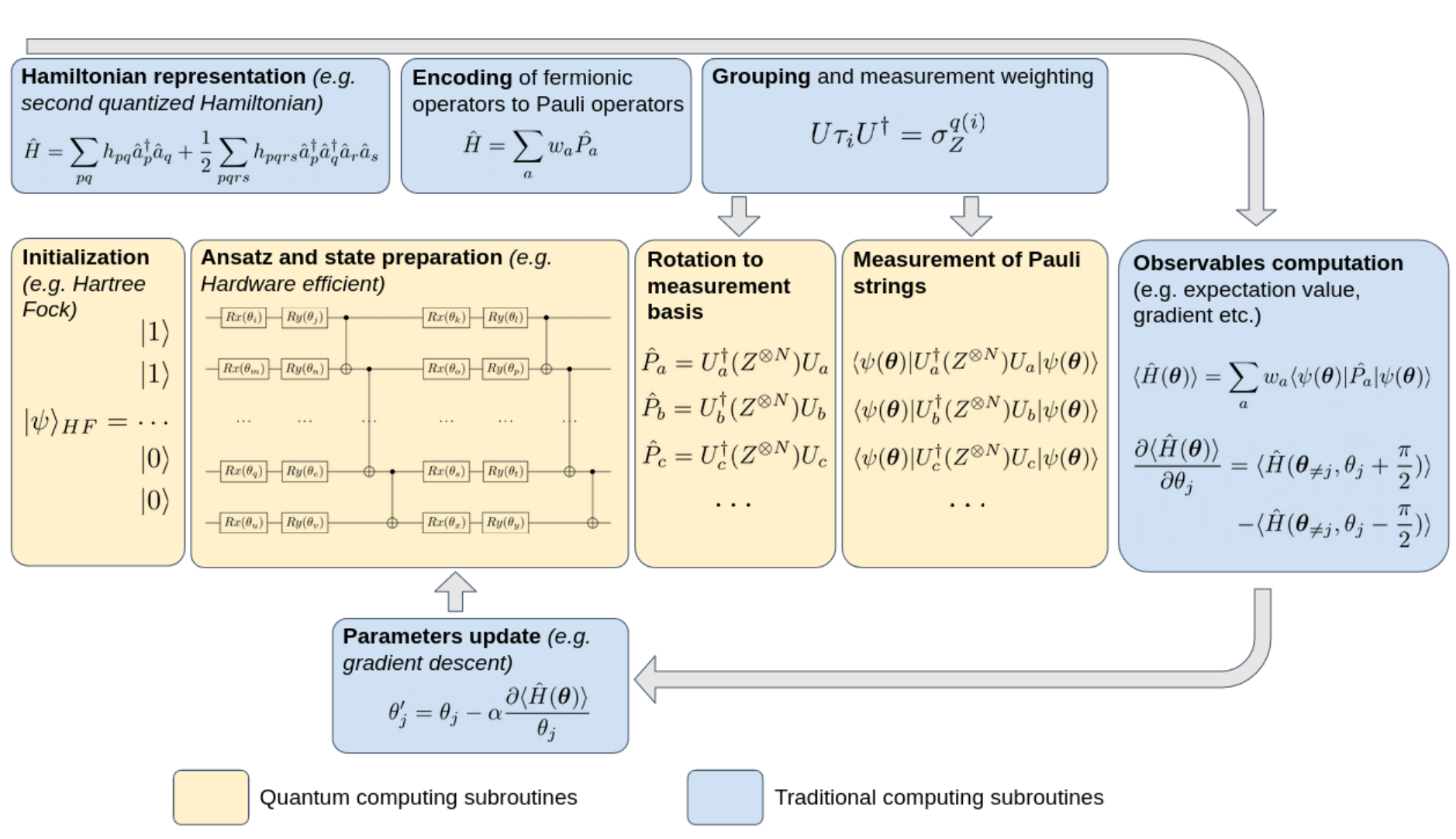}
    \caption{A complete plan of the VQE stack \cite{vqerev} }
    \label{fig:vqestack}
\end{figure}

Any Hamiltonian $\Ham$ has a lower eigenvalue $E_G$, the so-called \textit{ground state energy}.
So the variational method stands that for any (normalized) state $\ket{\psi}$, the following relation is verified :
\begin{equation}
    E_G \leq \braket{\psi | \Ham | \psi}
\end{equation}
One can then minimize the expression $\braket{\psi | \Ham | \psi}$ over all the states $\ket{\psi}$ in order to find a value closed to the ground state energy (see the complete VQE process in Fig \ref{fig:vqestack}).

In the VQE algorithm, generating those states is made by the \textit{ansatz}. Finding the good ansatz which generate an interesting bunch of state is an hard-problem due to the properties of the Hilbert space. An ansatz formed by a polynomial number of gate will only generate a polynomial number of states.
\subsection{Hardware Efficient Ansatz}\label{Hardware efficient ansatz}
As seen before, to approximate the state $\ket{\psi_{G}}$ corresponding to the ground state energy $E_G$, the problem of building the right ansatz is the hard part. Thus, when constructing such an ansatz we must balance two opposing goals. 

Since our state $\ket{\psi_{G}}$ belongs to a specific Hilbert space $\mathscr{H}$, ideally we would like our $n$ qubits ansatz to \textbf{span as much states} of $\mathscr{H}$ as possible, i.e generate any possible state $\ket{\psi}$ where $\ket{\psi}\in\mathbb{C}^N$ and $N=2^n$ ($n$ the number of qubits available). This is called the \textit{expressibility} of the ansatz \cite{expre}. However we would also like to use as few parameters as possible and so as few to optimise. The main idea of the Hardware efficient ansatz (HEA) is to focus on the first goal. 

\subsubsection{General construction}
\begin{figure}[H]
    \centering
    \includegraphics[scale=0.7]{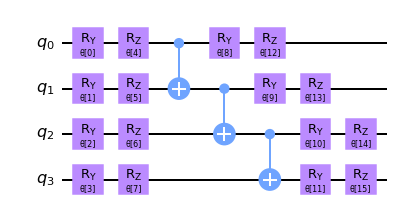}
    \caption{Example of one repetition of HEA in Qiskit, using $R_y$, $R_z$ and $CNOT$ gates, on four qubits.}
    \label{fig:simphea}
\end{figure}

Indeed the initial motivation for designing the HEA was for the trial state to be parametrized by quantum gates directly tailored to the quantum device available, and so spanning as much states available to the device. Despite depending on the device used, every HEA follows the same logic: the ansatz is constructed by repeating blocks of inter-weaved single qubit parametrized rotation gates and ladders of entangling gates \cite{vqerev} (see Fig. \ref{fig:simphea}). This can be represented as follow :
\begin{equation}
    \ket{\psi(\theta)}= \left[\prod_{i=1}^{d} U_{rota}(\theta_i)\times U_{ent}\right]\times U_{rota}(\theta_{d+1})\ket{\psi_{init}}
\end{equation}

where $d$ is the number of repetition of the scheme, $U_{rota}(\theta_i)=\prod_{q,p}^{}R_{p}^q(\theta_i^{pq})$, $q=0,1,...,n$ the qubit addresses and $p\in\{X,Y,Z\}$, the selected set of parametrized Pauli rotation.

\begin{figure}[H]
    \centering
    \includegraphics[scale=0.7]{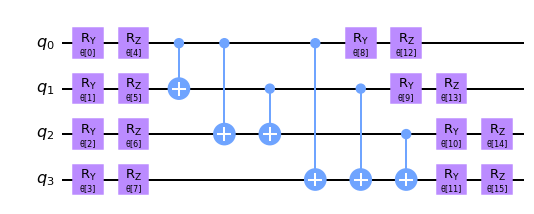}
    \caption{Same as figure \ref{fig:simphea} but with \textit{full entanglement}}
    \label{fig:simpheafull}
\end{figure}

HEA can be constructed differently depending on the depth of the circuit, the type of rotation gates used and the entanglement setting. This last parameter set the level of entanglement of the circuit, we show here two types of entanglement settings, \textit{linear} where CX gates are applied to adjacent qubits pairs in order, as shown in figure \ref{fig:simphea}, and \textit{full entanglement}  where a CX gate is applied to each qubits pair in each layer, as shown in figure \ref{fig:simpheafull}.

This way, good approximations on a wide range of states may be obtained despite a number of parameters in $\mathcal{O}(nd)$. Its versatility and ease of construction resulted in it being widely used for numerous small-scale quantum experiments.

\subsection{Unitary-Coupled Cluster Ansatz}

\subsubsection{Mean-field method or self-consistent field method (Hartree-Fock)} \label{Hartree-Fock}

The solution of the self-consistent field method, the so-called Hartree-Fock state, is an approximation of the wave function under the form of single Slater determinant. From the variational method, one can derive a set of N equation for the N spin-orbitals. This equation is solved by means of an iterative method.
In the Mean-field method the effect of all the other individuals on any given individual is approximated as a single effect, reducing the many-body problem to a one-body problem. Finding the one-electron HF wavefunction is equivalent to solve the following equation:
\begin{equation}
    \hat{F}\phi_i(1) = \epsilon_i \phi_i(1) 
\end{equation}
where $\hat{F}$ is the Fock operator and $\phi_i(1)$ are a set of one-electron wave functions, the HF molecular orbitals.

In the molecular orbital basis the HF state can be written as: $\ket{HF} = \crea_n\crea_{n-1}\ldots\crea_1\ket{vac}$ while in the spin-orbital basis it is written as : $\ket{HF} = (\crea_1)^{\phi_1}(\crea_2)^{\phi_2}\ldots(\crea_n)^{\phi_n} \ket{vac}$.

Hartree-Fock method is unable to approximate the electron correlation effect close to chemical accuracy.
To solve the problem above, one can expand the wavefunction into the $n$-Fock space as a superposition of all determinants. The coefficients of the superposition can be parametrized via two popular methods which are Configuration Interaction (CI) and Coupled-Cluster (CC) \cite{ucc}.

\subsubsection{Configuration Interaction (CI)}

The full configuration interaction (FCI) expands the wavefunction in the $n$-Fock space but become quickly intractable due to the number of determinants which is factorial.
One can limit this approach to only determinants with a fixed number of excitations from a reference state which is often chosen as the $\ket{HF}$ state : this is the CI method. Excitation operators are defined as :
\begin{align}
    \exci &= \sum_{i=1}^n \exci_i\\
    \exci_1 &= \sum_{\substack{i \in occ\\ a \in virt}} t_a^i \crea_a \anhi_i\\
    \exci_2 &= \sum_{\substack{i>j \in occ\\ a>b \in virt}} t_{ab}^{ij} \crea_a \crea_b \anhi_i \anhi_j \\
    \cdots
\end{align}

where $occ$ and $virt$ space are the occupied and unoccupied sites in the reference state. One cans see the operator $\exci_1$ generate one excitation for one electron of the reference state, the operator $\exci_2$ two excitation and so on. $t_a^i$ and $t_{ab}^{ij}$ correspond to expansion coefficients to minimize. In the exact FCI one has :
\begin{equation}
    \ket{FCI} = (1+\exci)\ket{HF}
\end{equation}
FCI solution can always be approached by increasing the number of excitation operators. The most common used CI is the CISD (Configuration Interaction Single Double) which contains only the terms $\exci_1$ and $\exci_2$.

There are several problems for the Configuration Interaction method :
\begin{enumerate}
    \item The method converge slowly for highly correlated states
    \item It is not size-extensive
\end{enumerate}

\subsubsection{Coupled Cluster}

The problems of the CI can be solved partially by the Coupled Cluster (CC) using an exponential ansatzs :

\begin{equation}
    \ket{\psi} = e^{\exci} \ket{HF}
\end{equation}

In this scenario, the parameters $\Vec{t}$ constitute the excitation amplitudes instead of the expansion coefficients. As for CI, CC is truncated for some fixed level, for instance the CCSD :
\begin{equation}
    \ket{\psi} = e^{\exci_1 + \exci_2} \ket{HF}
\end{equation}

A major difference between CI and CC is that, for a same level of truncation, CI involve only a polynomial number of determinants while CC involve all determinants in the $n$-Fock space.

Coupled Cluster also have some problems :
\begin{enumerate}
    \item It converges only under the assumption of a single reference state, mutli-reference is not possible using this method.
    \item The operator $e^{\exci}$ are not unitary ! This implies that it is not possible to use the variational principle.
\end{enumerate}

\subsubsection{Unitary Coupled Cluster}

The Unitary Coupled Cluster come to solve the problems of CC.
We redefine the operators to be unitary \cite{ucc} :
\begin{equation}
\label{eq:ucc}
    \ket{\psi} = e^{\exci - \exci^\dagger} \ket{HF}
\end{equation}

One can know use the variational principle as follow :
\begin{equation}
    E_G = \min_{\Vec{t}} \braket{HF|e^{-(\exci - \exci^\dagger)}\Ham e^{\exci - \exci^\dagger}|HF}
\end{equation}

Another point to notice is that the space span by this ansatz is the same as the one span by the CC.

We know may want to implement the UCC using a Quantum Computer. This ansatz require 2 steps : \begin{enumerate}
    \item Preparing the reference state $\ket{HF}$. One just need to run the Hartree-Fock Algorithm to obtain a good reference with a significant overlap.
    \item Application of the UCC unitary $U(\Vec{t})$ in order to prepare a wavefunction to evaluate.
\end{enumerate}

Let's focus on the second step, to create such an Unitary one need to map the UCC operators \ref{eq:ucc} to operations that can be performed on a quantum computer using some map (for example Jordan-Wigner). First we need to re-write the operator a little bit :
\begin{equation}
    U(\Vec{t}) = e^{\sum_j t_j(\sexci_j - \sexci_j^\dagger)}
\end{equation}
with $\sexci_j$ the excitations  $\crea_a\anhi_i$ , $\crea_a\crea_b\anhi_i\anhi_k$\ldots and $t_j$ the CC amplitudes.
Because excitation operators do not commute, one need to use \textit{trotterization} \cite{trotter} (more or less to separate the exponential of a sum of non-commutable operators into products of exponentials) to approximate and get for trotter coefficient equal to 1:
\begin{equation}
    U_1(\Vec{t}) = \prod_j e^{t_j(\sexci_j - \sexci_j^\dagger)}
\end{equation}
Applying the Jordan-Wigner transformation one get :
\begin{equation}
    \label{eq:opUCC}
    (\sexci_j - \sexci_j^\dagger) = i \sum_k^{2^{2t_j - 1}} P_k^j
\end{equation}
where the $P_k^j$ are some products of Pauli matrices referred as sub-terms.
For instance the single cluster excitation terms can be written as :
\begin{equation}
    t_i^a(\crea_a\anhi_i - h.c.) = \frac{it_i^a}{2} \bigotimes_{k=i+1}^{a-1} Z_k(Y_i X_a - X_i Y_a)
\end{equation}
The transformed UCC operator is then :
\begin{align}
    U_1(\Vec{t}) &= \prod_j \exp\Bigg(i t_j \sum_k^{2^{2t_j -1}} P_k^j\Bigg)\\
    \label{eq:uccq}
    &= \prod_j \prod_k^{2^{2t_k-1}}\exp(i t_j P_k^j)
\end{align}
second equation is obtained using commutation of the sub-terms derived from same excitation.
Equation \ref{eq:uccq} can be implemented on a quantum computer using qubits and the regular model of quantum computing.

The most popular version of UCC is UCCSD where one keep only the Single and Double excitation:
\begin{equation}
    \exci \approx \exci_1 + \exci_2
\end{equation}

The major problem of UCC ansatz lies in the number of parameters to optimize evolving quartically with the number of spin-orbitals.

\subsection{ADAPT-VQE}\label{ADAPT-VQE}

The ADAPT-VQE is a VQE method that has the particularity of building the ansatz during the optimisation process. It starts by selecting a pool of operators and adds them progressively to the circuit by choosing at each step the operator that lowers the most the energy of the currently constructed state. Different ADAPT-VQE methods appear by choosing different pools of operators.

\subsubsection{The Fermionic-ADAPT-VQE}

In the Fermionic ADAPT-VQE \cite{grimsley_adaptive_2019} the operators of the pool are the ones that appear in the UCC ansatz (\ref{eq:opUCC}). The algorithm starts from a state $\ket{\psi_0}$ (which is often the Hartree-Fock state) and initialize the ansatz to the identity. It then iteratively construct the states $\ket{\psi_k} = e^{\theta_k^kA_k}...e^{\theta_1^kA_1}\ket{\psi_0}$. To do so it determines at each step the operator from the pool that lowers the most the energy of the current state, ie. the operator $e^{\theta_n A_n}$ with $\theta_n$ initialized to $0$ such that 

\begin{align}
 \frac{\partial}{\partial \theta_n} \left. \bra{\psi_k} e^{-\theta_n A_n} H e^{\theta_n A_n} \ket{\psi_k}\right|_{\theta_n = 0} & = \left. \bra{\psi_k} -Ae^{-\theta_n A_n} H e^{\theta_n A_n} + e^{-\theta_n A_n} H A e^{\theta_n A_n} \ket{\psi_k}\right|_{\theta_n = 0} \\
& = \bra{\psi_k} H A - A H \ket{\psi_k} \\
& = \bra{\psi_k} [H,A] \ket{\psi_k}
\end{align}

is maximal. Once the new operator has been chosen, a full VQE optimization is run on all the parameters $\theta_1,...,\theta_k$ to obtain the lowest energy for a state with the current ansatz. The process stops when adding a new operator to the ansatz has a negligible impact on the energy of the state (ie. when $\forall n,\frac{\partial}{\partial \theta_n} \left. \bra{\psi_k} e^{-\theta_n A_n} H e^{\theta_n A_n} \ket{\psi_k}\right|_{\theta_n = 0} = \bra{\psi_k} [H,A] \ket{\psi_k} \leq \epsilon$) where $\epsilon$ is a predefined parameter.

The whole process of the Fermionic-ADAPT-VQE can also be seen as trying to approximate the Full Configuration Interraction (FCI)

\begin{equation}
    \ket{\psi^{FCI}} = \prod_0^{\infty}\prod_{n=1}^{N^4}e^{\theta_n^kA_n}\ket{HF}
\end{equation}

but with the constraint that at each step only one null parameter can be changed.

\subsubsection{The Qubit-ADAPT-VQE}

The Qubit-ADAPT-VQE \cite{tang_qubit-adapt-vqe_2021} follows the exact same steps as the Fermionic-ADAPT-VQE but the operators constructing the ansatz are chosen of the form

\begin{equation}
    A = i \bigotimes_{k=1}^N P_k
\end{equation}

where $P_i$ is an operator of the Pauli group. The pool can in fact be reduced by only taking the Pauli strings that appear in the fermionic excitation mapping that verify time-reversal symmetry and removing the Pauli $Z$ operators used for anti-symmetry of the wave function in the Jordan-Wigner mapping since they do not impact the energy estimate.
Compared to the Fermionic-ADAPT-VQE this method reduces the number of C-NOT gates in the ansatz but at the cost of a wider pool of operators and more parameters to optimize to achieve a wanted precision.

\subsubsection{The QEB-ADAPT-VQE}

The Qubit-Excitation-Based(QEB)-ADAPT-VQE presented in \cite{yordanov_qubit-excitation-based_2021} is an ADAPT-VQE method that is very similar to the two previous algorithms. However it has two differences with the Ferminionic-ADAPT-VQE:
\begin{itemize}
    \item The pool of chosen operator is different. The one- and two-body operators from the Fermionic-ADAPT-VQE are replaced by single- and double-qubit-excitation which are generated by the following creation and annihilation operators:
    \begin{equation}
        Q_j^{\dagger} = \frac{X_j -iY_j}{2} \quad \quad \quad \quad  Q_j = \frac{X_j +iY_j}{2}
        \label{QI}
    \end{equation}
    ie. operators that are defined as the Fermionic operators in the Jordan-Wigner mapping but without the Pauli-$Z$ strings that account for the anti-commutation relations.
    One can then define the skew-hermitian single- and double-qubit-excitation operators
    \begin{equation}
        T_i^k(\theta) = \theta(Q_k^{\dagger}Q_i - Q_i^{\dagger}Q_k) \quad \quad T_{ij}^{kl}(\theta) = \theta(Q_k^{\dagger}Q_l^{\dagger}Q_iQ_j-Q_i^{\dagger}Q_j^{\dagger}Q_kQ_l) 
        \label{TI}
    \end{equation}
    which are represented on the ansatz by the unitaries
    \begin{equation}
        U_{ki}(\theta) = e^{T_i^k(\theta)} \quad \quad U_{klij}(\theta) = e^{T_{ij}^{kl}(\theta)}
        \label{ETI}
    \end{equation}
    \item At step $n$, when it determines the next operator to add to the ansatz, instead of choosing the one with with highest $\frac{\partial}{\partial \theta_i} \left. \bra{\psi_n} e^{-\theta_i A_i} H e^{\theta_i A_i} \ket{\psi_n}\right|_{\theta_i = 0}$ it keeps the top $k$ operators (with $k$ a predefined number). It then runs a complete VQE for each of the $k$ new ansatzes obtained by adding the different operators to the current circuit and finaly picks the operator that gave the lowest energy.
\end{itemize}
The main interest of this method is that it has less CNOT gates than the qubit-ADAPT-VQE but the same amount of parameters as the Fermionic one.

\section{Other aspects of VQE}
\subsection{Classical Optimisers} 
\label{Optimizers}

Hybrid algorithms like the VQE allegedly uses the best of both worlds, classical optimisation and quantum mechanics. Therefore in order to have a well functioning VQE one also need a sufficient optimiser. There is a plethora of different classical optimisers that are available and implemented within the Qiskit environment. Qiskit suggest that when running a VQE on NISQ hardware one should use the classical optimiser, Simultaneous Perturbation Stochastic Approximation (SPSA) \cite{noauthor_simulating_nodate}. Otherwise a common optimiser when running on a classical quantum simulator is to to use Constrained Optimisation By Linear Approximation (COBYLA).

Lavrijsen et. all showed in \cite{lavrijsen_classical_2020} that in order to have a well functioning VQE on quantum noise hardware the classical optimiser must be carefully chosen as well. Furthermore, in order to get valid scientific results on NISQ hardware they argue that choosing a well tuned optimiser is crucial. In their paper they test how different optimisers behave when applying noise to the quantum gates in a classical quantum simulator. The results show that classical optimisers who are unable to take noise in consideration end up loosing chemical accuracy for a lower amount of noise than the optimisers that are noise-aware. Best performing optimiser under the presence of noise for their quantum chemistry example was the Implicit Filtering (ImFil) \cite{noauthor_4_nodate}.  ImFil is implemented as in MATLAB but as part of the work with the article Lavrijsen et. all has rewritten the original MATLAB implementation to Python. Which can be acessed by the SCIKIT-QUANT library that consists of all the optimisers they have analysed in the paper \cite{noauthor_qat4chem_nodate}.


\subsubsection{Sequential Least Squares Programming (SLSQP)}

Qiskit uses the Scipy package scipy.optimize.minimize SLSQP \cite{noauthor_scipyoptimizeminimize_nodate}. Which uses the Sequential Least Squares Programming method. That was developed by Dieter Kreft in 1980s. The algorithm is used to optimise general nonlinear problems that minimize a scalar function.

\subsubsection{Constrained Optimisation By Linear Approximation (COBYLA)}

Qiskit uses the Scipy package scipy.optimize.minimize COBYLA. Which uses the Constrained Optimisation By Linear Approximation method. That was developed by Powell in 1994 \cite{gomez_direct_1994}. The algorithm optimises a constrained problem where the function has no derivative. It works by each iteration forms a linear approximation to the objective and constraint function by interpolation at the vertices of a simplex and trust region bound restricts each change to the variable.

\subsubsection{Simultaneous Perturbation Stochastic Approximation (SPSA)}
Qiskit uses the Scipy package  scipy.optimize.minimize SPSA.The SPSA approximates the gradient of the objective function using two measurements per iteration \cite{noauthor_spsa_nodate}. This is regardless of the dimension of the optimisation problem. The two measurements are made by randomly varying the all of the variables in the problem so called simultaneous perturbation. Comparing with the finite-difference method that need to make $2p$ measurements per iteration when optimising $p$ parameters, varying one variable at the time.

\subsubsection{Implicit Filtering (IMFIL)}
Qiskit uses the Scikit package where Lavrijsen et. all rewrote the MATLAB implementation to python. The IMFIL algorithm is designed for problems with local minimas that is caused by noise. IMFIL is a sampling method which means that the optimisation is controlled by evaluating the function $f$ at a cluster of points. Depending on the local minima found by the cluster through interpolation drives the next cluster of points. The optimiser converges when the cluster has reached it's smallest size.

\subsection{Barren Plateau}

In this section, we will introduce the problem of Barren Plateau (BP) which is common in VQA. BP refers to an issue in the choice of the cost function (which for us is the expectation value of the Hamiltonian on the current state) or of the optimizer. When a BP of the cost function is reached by the optimizer, the gradient will vanished in many/all directions causing a lot of troubles for the optimizer to reach the global minimum. It has the effect of the landscape being essentially flat, for instance when reaching a local minima. Then depending on the existence or not of BP, one can choose to use a gradient-based or a gradient-free optimization method. It was shown that deep unstructured parametrized quantum circuits tend to exhibit BPs when they are randomly initialized.

\chapter{Implementation}

\paragraph{Implementation link:}Links for the notebooks are in reference \cite{nb_vqe_adapt}, \cite{nb_uccsd_HEA}, \cite{nb_uccsd_h2}, \cite{nb_uccsd_LiH}. 
    
We have chosen to implement the VQE algorithm using the open-source framework Qiskit Nature \cite{noauthor_qiskit_2022} on IBMQ's quantum computer and classical simulator. The primary reason for choosing Qiskit as our software development kit over OpenFermion (Google's open source library) is its large community of users and its ease of use. Additionally the ADAPT-VQE algorithm was first implemented using OpenFermion in \cite{grimsley_adaptive_2019} and it seemed more interesting to try to do it with another development kit.

\begin{figure}[H]
    \centering
    \includegraphics[scale=0.3]{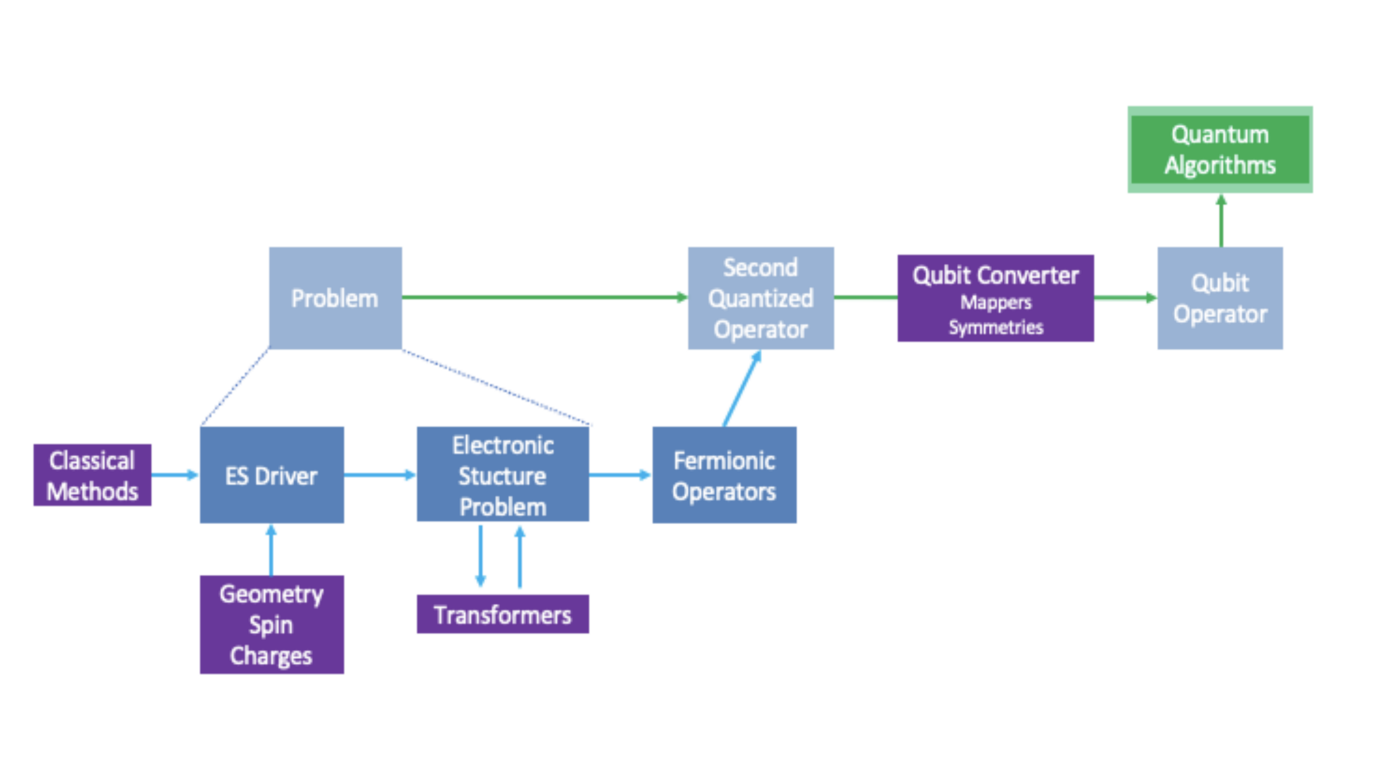}
    \caption{Qiskit Nature outline, \cite{qiskit_introducing_2021}}
    \label{fig:qiskit_nature}
\end{figure}

In fig(\ref{fig:qiskit_nature}) we see the structure of Qiskit Nature and the different parts that are necessary to finally run a quantum algorithm, in our case a VQE.

\section{Qiskit general framework}
\subsection{Hamiltonian Mapping}
As the goal of our implementation is to find the ground state energy of a particular molecule, we start by describing the energy of our system with the following Hamiltonian
\begin{equation}
    \Ham = - \left(\sum_i \frac{\hbar^2}{2m_e} \nabla_i^2 +  \sum_{i,k} +  \frac{e^2}{4\pi\epsilon_0}\frac{Z_k}{|r_i-R_k|} +
    \frac12 \sum_{i \neq j} \frac{e^2}{4\pi\epsilon_0}\frac{1}{|r_i-r_j|}\right).
\end{equation}
We rewrite the Hamiltonian in the second quantisation form (see equation (\ref{second_quant})) using the molecular orbitals (MOs), $\phi_u$ together with the annihilation and creation operators (see \nameref{First and Second Quantization of the Hamiltonian})
\begin{align}\label{second_quant}
    &\Ham = \sum_{pq} h_{pq}\crea_p\anhi_q + \frac12 \sum_{pqrs} h_{pqrs}\crea_p\crea_q\anhi_r\anhi_s \\
    & h_{pq}  =  \int d\xi \phi_p^*(\xi) \Big(-\frac{\hbar^2}{2m_e}\nabla^2 - \sum_k \frac{e^2}{4\pi\epsilon_0}\frac{Z_k}{|r-R_k|}\Big)\phi_q(\xi) \\
    & h_{pqrs}  =  \frac{e^2}{4\pi\epsilon_0}\int d\xi_1d\xi_2 \frac{\phi_p^*(\xi_1)\phi_q^*(\xi_2)\phi_r(\xi_1)\phi_s(\xi_2)}{|r_1 - r_2|}
\end{align}
The MOs are found by solving the Hartree-Fock (HF) state as described earlier in \nameref{Hartree-Fock}. This is done by using the classical solver PySCF \textit{driver} \cite{noauthor_home_nodate} which is an open-source collection of electronic structure modules. To get the second quantized Hamiltonian we use the function ElectronicStructureProblem.second\_q\_op \cite{noauthor_electronicstructureproblem_nodate} on the MOs which then returns the operators for the Hamiltonian.

We now have the Hamiltonian of the initial HF state and need to rewrite it in a way a quantum computer can handle. As each qubit in a quantum computer can be used to represent a spin orbital we once again rewrite the Hamiltonian using the Jordan-Wigner mapping (\nameref{Mapping}) which assigns each qubit to a spin orbital by mapping the creation operators as in equation (\ref{eq:jordan-wigner}).

\subsection{Operators representation}\label{Operators} 

To represent unitaries (and in general the operators we handle during the various implementation) we use different kind of classes. To decompose pre-existing ansatzes (such as UCCSD), we use  the \texttt{Instruction} class\cite{instruction_qiskit} which represent circuit instructions, which in our case are excitation operators. \\
When we create operators from scratch, we obviously have to start with Paulis strings, then we can use ether the \texttt{PauliOp} class \cite{pauliop_qiskit}, which can only represent a tensor of Paulis, or the \texttt{SparsePauliOp}\cite{sparsepauliop_qiskit} that can represent a linear combination of tensor of Paulis. Then when we want to create unitaries that represent parameterized excitation operator as in \ref{ETI}, we have to call \texttt{PauliEvolutionGate}\cite{hamiltoniangate_qiskit} on the two class mentioned above, which can represent such unitaries, and then be appended to a circuit.

\subsection{Aer simulator} 
To simulate our circuits/ansatzes, we use the \texttt{AerSimulator} backend \cite{aer_qiskit} from IBM Qiskit. The default behavior of this simulator is to mimic the execution of an actual noise-free device. Obviously the major drawback of this technology is the computation time which starts being very long when we compute it for molecules different than $H_2$.

\subsection{VQE}
To run a VQE, we use the qiskit class VQE \cite{noauthor_vqe_nodate}. This class takes an ansatzes, an optimizer, an initial state (usually the HartreeFock state \ref{Hartree-Fock}), a quantum instance such as the \texttt{AerSimulator}\cite{aer_qiskit}, and an initial point representing the starting point for the vector of parameters. We then call the \texttt{compute\_minimum\_eigenvalue()} method which starts the whole VQE process and returns a data type with mutliple interesting values (optimizer time, number of cost function evaluation, eigenvalue, eigenstate ...).

\section{Pre-existing Ansatz}
\subsection{Hardware Efficient}
As mentioned previously in \nameref{Hardware efficient ansatz} the ansatz is made of a series of repeating rotation gates followed by CNOT gates. This is implemented using the EfficientSU2 function \cite{noauthor_efficientsu2_nodate} that creates a circuit with the specific pattern of gates described before. 

\subsection{UCCSD}\label{UCCSD}
For the UCCSD ansatz we only keep the first and second excitation operators as discussed previously. These are generated by using the generate\_fermionic\_excitations function \cite{noauthor_generate_fermionic_excitations_nodate}.

\section{ADAPT-VQE development}
\begin{figure}[H]
    \centering
    \includegraphics[width=\textwidth]{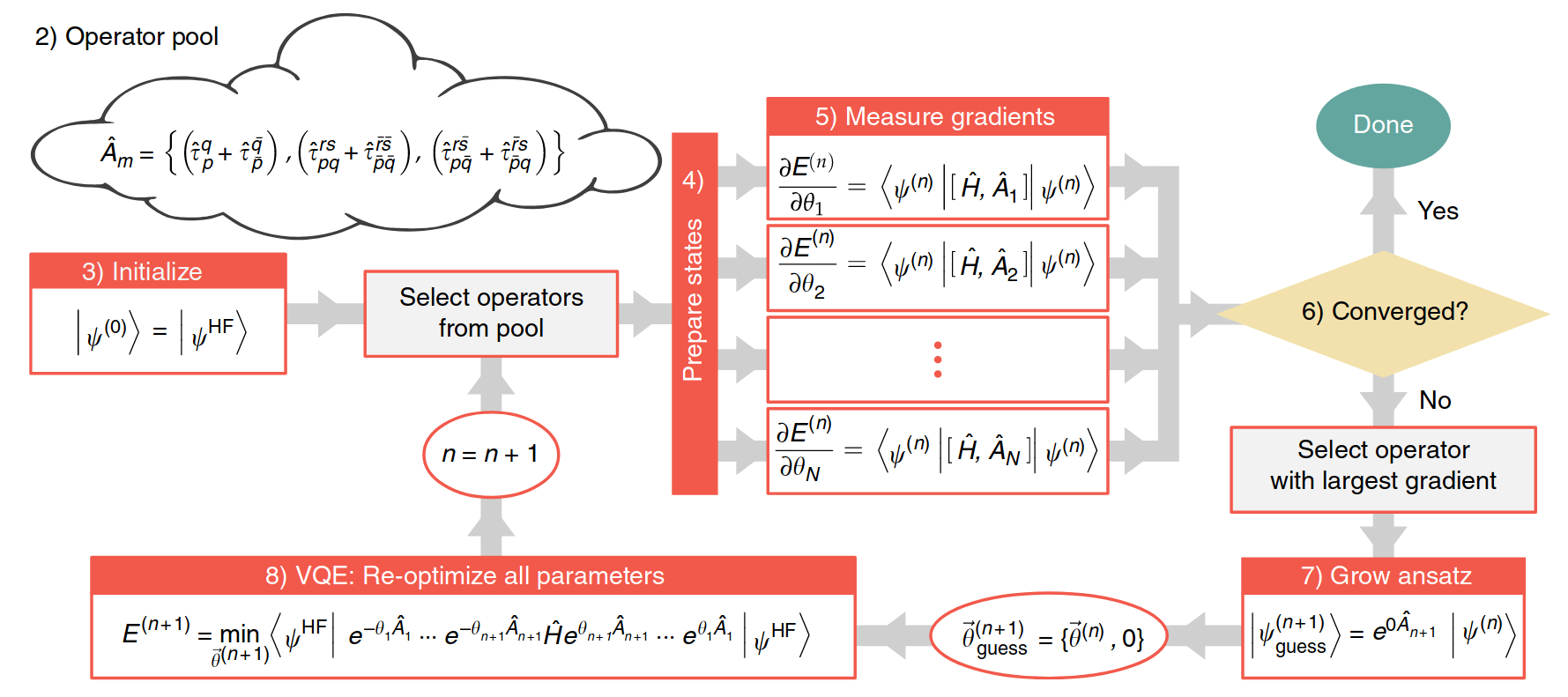}
    \caption{Adapt-VQE schematic, \cite{grimsley_adaptive_2019}}
    \label{fig:adapt_scheme}
\end{figure}

This part concerns all the development done to implement the three ADAPT-VQE methods introduced in \nameref{ADAPT-VQE}. There exist an AdaptVQE class \cite{noauthor_adaptvqe_nodate} in Qiskit wich aims to implement the Fermionic-ADAPT-VQE method, but due to several \texttt{raise NotImplementedError} exceptions in the subclasses and a unexplicit description of the method, we have decided to start from scratch.
We follow the scheme in fig(\ref{fig:adapt_scheme}). The main focus in the development of these methods, is the definition of the pool of operators.

\subsection{Fermionic-ADAPT-VQE}
  After initializing our state to the Hartree-Fock state as for the \nameref{UCCSD}, we start with an empty ansatz and measure which operator from the pool achieves the largest gradient using the \texttt{commutator(op\_a, op\_b)}\cite{commutator_qiskit} class between the operators and the hamiltonian and then append that operator to the ansatz (which is the case for every methods).\\
  In this case building the operator pool is pretty straightforward: since the Fermionic-ADAPT-VQE method use directly the excitation operators of the UCCSD ansatz, we construct one time this ansatz and then decompose every excitation operator to create our pool (cf. \nameref{Operators} to see each decomposition step of the ansatz using several Qiskit methods).  

\subsection{Qubit-ADAPT-VQE}
Here the operators considered are all possible combinations of the N-tensor of Pauli operators, N being the number of spin orbitals/qubits of our circuit. The numbers of operator can be widely reduced using only the Pauli strings which appear in the operators of the fermionic pool created before. The fermionic pool was represented as a list of \texttt{PauliSumOp}\cite{paulisumop_qiskit} which are made of \texttt{SparsePauliOp}\cite{sparsepauliop_qiskit} that contain a list of strings representing the Pauli strings. We can then extract the strings from the pool and, after removing the redundant elements, create a new pool with the associated \texttt{PauliOp}\cite{pauliop_qiskit} objects. The last step consists in reducing one more time the number of operators by removing randomly $3/4$ of them. This can be done without a significant loss of accuracy as it was shown in the article \cite{tang_qubit-adapt-vqe_2021}.

\subsection{QEB-ADAPT-VQE}
This method lies in between rigorous problem-tailored methods as for the fermionic one, and more machine learning-like methods as for the qubit one, were the pool of operator is not tailored to the physics of the problem, but more aimed to have a wide range of options. Due to this \textit{"in-between"} aspect, this version was the hardest to implement.\\
 Since it's easier to work with unitaries defined on the whole circuit, to create the $Q_i$ described in \ref{QI}, we create the sum of string of paulis  $Q_i'= I^{\otimes i-1}\otimes Q_i \otimes I^{\otimes N-i-1}$ defined on the whole qubit space. The same way, we then define the  $T_i^{'k}$ as in \ref{TI} and finally the whole unitary $e^{T_i^{'k}}$ as in \ref{ETI} (following \nameref{Operators} section as always to create our circuit).

\chapter{Results and discussion}

The main goal of this chapter is to evaluate the accuracy and computing performance arising from the different ansatzes described above \ref{Ansatzes}. The framework of the experiments is sampling the potential energy surface (PES) of different molecules, $H_2$ and $LiH$. All of this is done in order to provide a better understanding of all the theoretical notion addressed in the previous sections.
\section{Comparison of various HEA}

In this section, we will compare different HEA constructions trough a simple experiment, computing the PES of $H_2$, since its scope of application is restricted to small sized quantum experiment. We will evaluate the effectiveness of the different construction methods with 2 criteria:
\begin{enumerate}
    \item The error in the calculated ground energy $E_{HEA}$ with respect to the FCI solution, $E_{FCI}$.
    \item The number of function evaluations required for convergence
\end{enumerate}
First lets display an overview of the different methods by testing them with different (\nameref{Optimizers} efficient ansatz) (COBYLA, L\_BFGS\_B, SLSQP,SPSA) in order to have a wider view of the ansatzes peformance, and to be sure as much as possible to not bottleneck it because of a unhadapted optimizer (we do not provide here a deep reflections about the compatibility of the optimizers, but only focus on the best ansatz performances) . The ansatzes considered will be: 
\begin{itemize}
    \item $R_yR_z$ with \textit{linear entanglement} 
    \item $R_yR_z$ with \textit{full entanglement}
    \item $R_x$  with \textit{linear entanglement} 
\end{itemize}
All of them with 3 repetition of the pattern (to keep the circuits shallow)

\begin{figure}[H]
\begin{subfigure}{.5\textwidth}
  \centering
  \includegraphics[width=.8\linewidth]{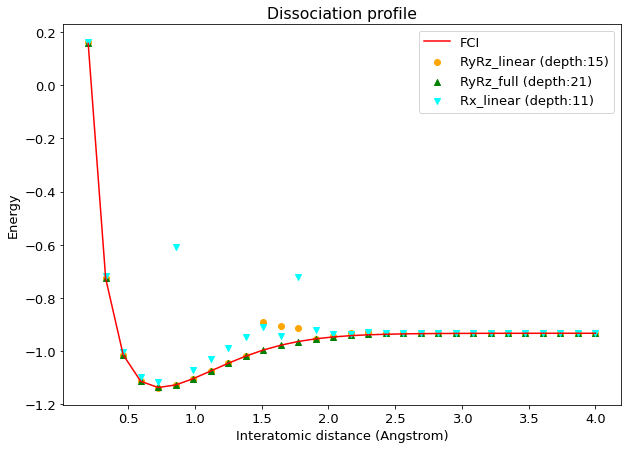}
  \caption{COBYLA}
  \label{fig:sfig1}
\end{subfigure}%
\begin{subfigure}{.5\textwidth}
  \centering
  \includegraphics[width=.8\linewidth]{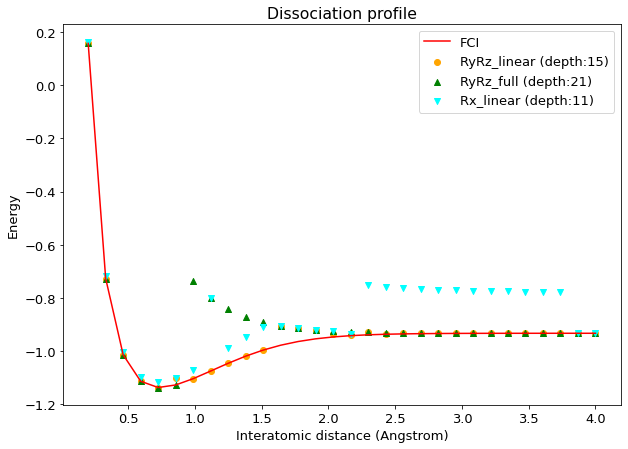}
  \caption{L\_BFGS\_B}
  \label{fig:sfig2}
\end{subfigure}
\begin{subfigure}{.5\textwidth}
  \centering
  \includegraphics[width=.8\linewidth]{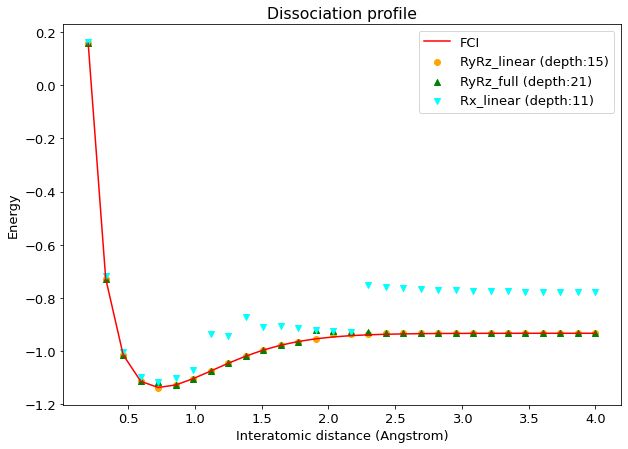}
  \caption{SLSQP}
  \label{fig:sfig3}
\end{subfigure}
\begin{subfigure}{.5\textwidth}
  \centering
  \includegraphics[width=.8\linewidth]{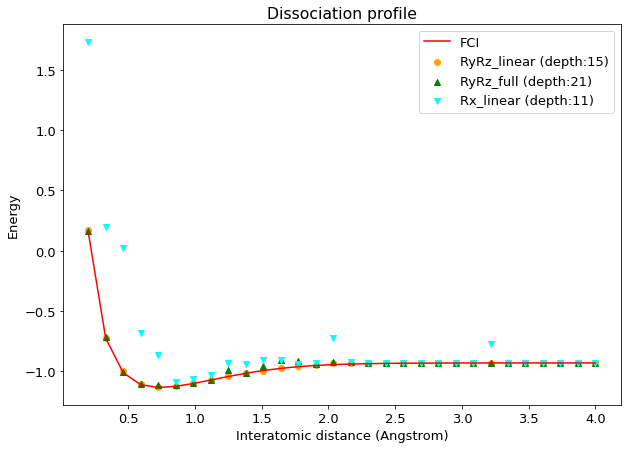}
  \caption{SPSA}
  \label{fig:sfig4}
\end{subfigure}
\caption{Energy calculated depending on the different optimizer used, for each HEA cited above with a threshold of 3000 evaluations.}
\label{fig:HEA_pes}
\end{figure}

 As we can see on Figure\ref{fig:HEA_pes}, $R_x$ having only one rotation gate, increases drastically the error compared to the FCI results. Indeed, the range of states it can access is reduced, having only one rotation axis per qubit, but with COBYLA \ref{fig:sfig1} $R_x$ was optimized well enough to tailor the FCI results for close ($[0.2,1]$) and far ($[2,4]$) inter-atomic distance . \\
 For the \textit{fully entangled} and \textit{linearly entangled} $R_yR_z$ the difference is not very noticeable (see \cite{PEScorr} for more reflections on entanglement-related aspects of the dissociation of $H_2$). However for the \textit{fully entangle} HEA in the case of \ref{fig:sfig2}, L\_BFGS\_B was not able to optimize well enough the circuit parameters resulting in a important error in the $[1,2]$ interval  . \\
 In general we get pretty good results for the two $R_yR_z$ HEA with \ref{fig:sfig1}, \ref{fig:sfig3} and \ref{fig:sfig4}.

\begin{figure}[H]
\begin{subfigure}{.55\textwidth}
  \centering
  \includegraphics[width=.8\linewidth]{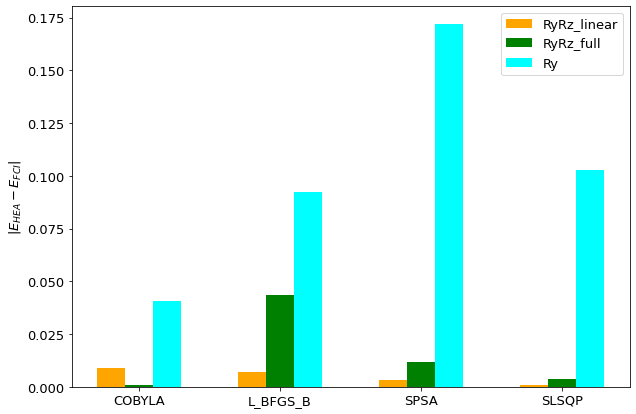}
  \caption{}
  \label{fig:sfig21}
\end{subfigure}%
\begin{subfigure}{.55\textwidth}
  \centering
  \includegraphics[width=.8\linewidth]{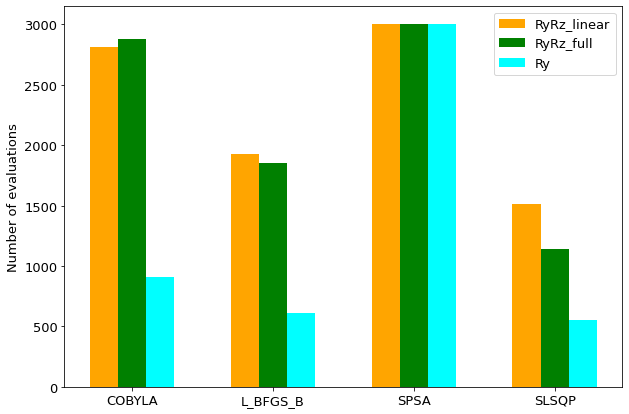}
  \caption{}
  \label{fig:sfig22}
\end{subfigure}
\caption{Overview of the error of the energy calculated with respect to the FCI solution and the number of function evaluations}
\label{fig:HEA_exp}
\end{figure}

The low number of function evaluations of the $R_x$ HEA on figure \ref{fig:sfig22}, confirm the fact that the simplest ansatz containing the solution will also have the best performance, thank to the solution space having less parameters to optimize, of course in this case, it is offset by the important errors brought by this ansatz for certain interatomic distances. We also see in \ref{fig:sfig21} the two $R_yR_z$ HEA having pretty small errors.

Overall despite not tailoring the problem considered, HEA ansatz can be interesting in the case of small chemical problems as the one shown above. With shallow circuit depth and few $CX$ gates it is possible to sample $H_2$ ground energy with good chemical accuracy. Since it has to span a large portion of the Hilbert space to represent with enough accuracy the ground state of the wave function considered, HEA can therefore be quite inefficient requiring in worst cases an exponential depth and so unlikely to be suitable for larger-scale chemical problems.

\section{Evaluation of the accuracy of UCCSD}
Now that we've define a lower bound for the performance of the ansatzes with HEA, let's evaluate our first problem-tailored ansatz, the UCCSD andatz. Firstly we will evaluate the performance of the UCCSD for the simulating the PES of $H_2$, hence to make sure that it is at least as accurate as HEA, and then we will do the same for $LiH$ and $BeH$, two more complex molecules, unreachable for reasonable HEA (we have actually computed simulation of these molecules for HEA, but the precision was to poor to be displayed or compared with other ansatzes).

\section{Comparison of ADAPT-VQE Methods}

\begin{figure}
\begin{subfigure}{.5\textwidth}
  \centering
  \includegraphics[width=0.8\linewidth]{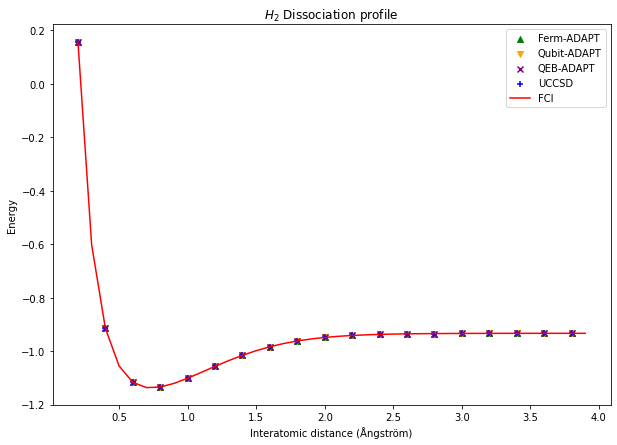}
  \subcaption{}
  \label{fig:H2PES}
\end{subfigure}
\begin{subfigure}{.5\textwidth}
  \centering
  \includegraphics[width=0.8\linewidth]{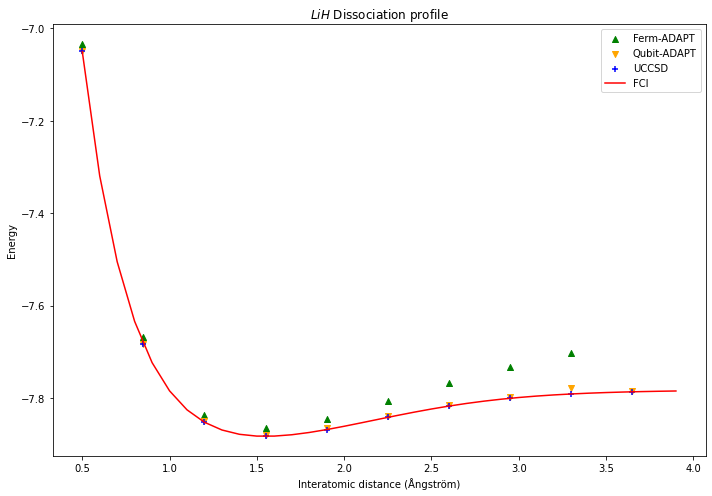}
  \subcaption{}
  \label{fig:LIHPES}
\end{subfigure}
\begin{subfigure}{.5\textwidth}
  \centering
  \includegraphics[width=0.8\linewidth]{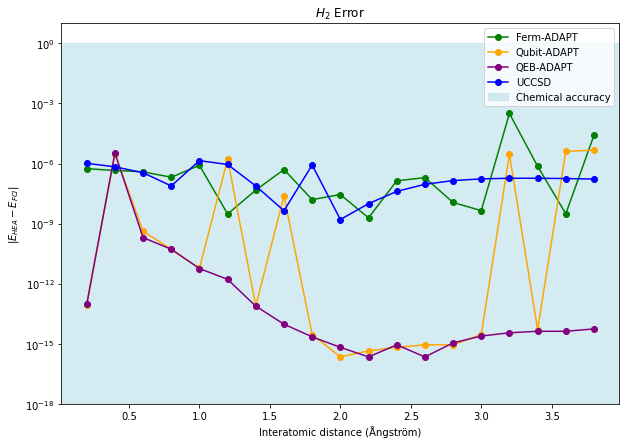}
  \subcaption{}
  \label{fig:H2ER}
\end{subfigure}
\begin{subfigure}{.5\textwidth}
  \centering
  \includegraphics[width=0.85\linewidth]{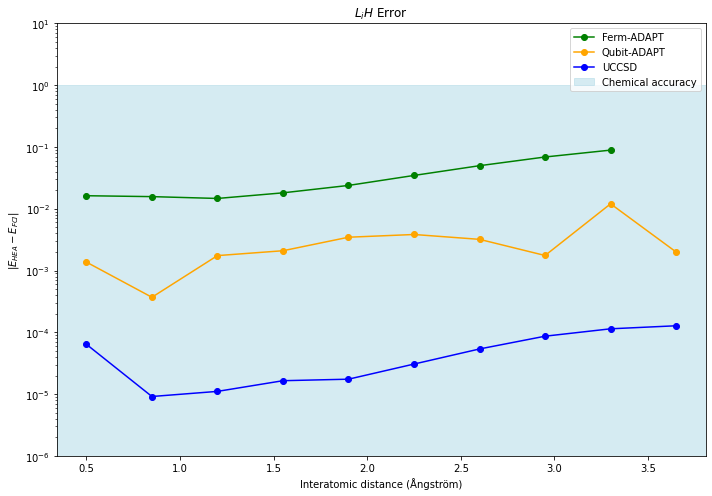}
  \subcaption{}
  \label{fig:LIHER}
\end{subfigure}
\begin{subfigure}{.5\textwidth}
  \centering
  \includegraphics[width=0.8\linewidth]{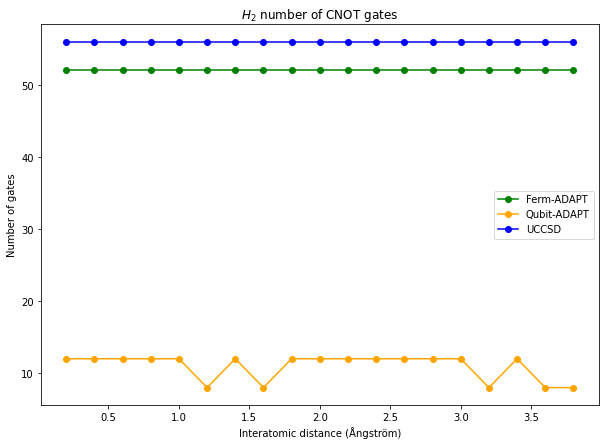}
  \subcaption{}
  \label{fig:H2CNOT}
\end{subfigure}
\begin{subfigure}{.5\textwidth}
  \centering
  \includegraphics[width=0.8\linewidth]{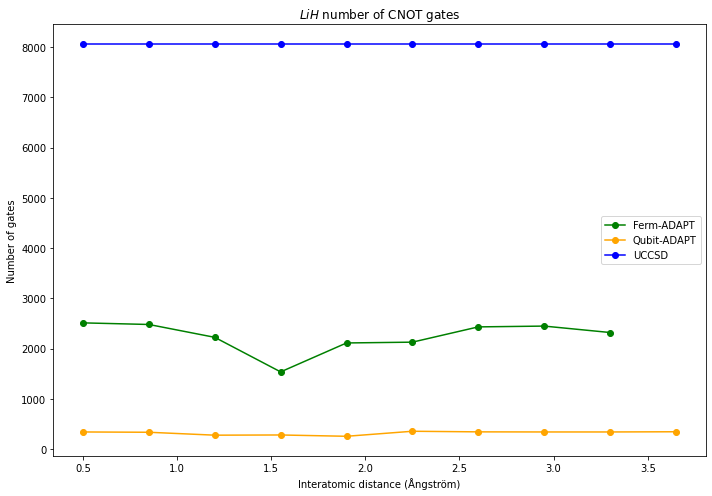}
  \subcaption{}
  \label{fig:LIHCNOT}
\end{subfigure}
\begin{subfigure}{.5\textwidth}
  \centering
  \includegraphics[width=0.8\linewidth]{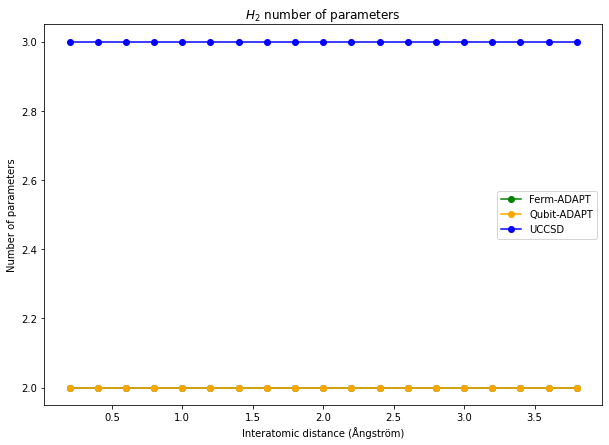}
  \subcaption{}
  \label{fig:H2PARAM}
  \caption*{$H_2$}
\end{subfigure}
\begin{subfigure}{.5\textwidth}
  \centering
  \includegraphics[width=0.8\linewidth]{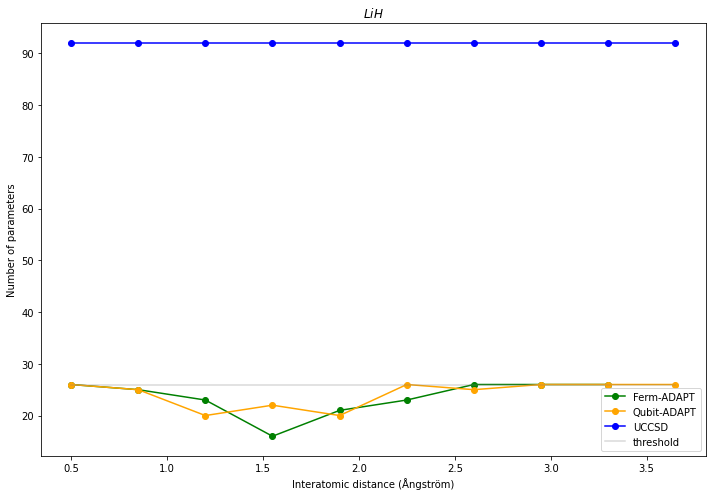}
  \subcaption{}
  \label{fig:LIHPARAM}
\caption*{$LiH$}

\end{subfigure}
\caption{}
\label{fig:HEA_pes}
\end{figure}

\subsection{Results for $H_2$}

\subsubsection{Accuracy}

On figure \ref{fig:H2ER}, one can see that each algorithm performs well with an error within the chemical accuracy (which is the accuracy required to make realistic chemical predictions).

Comparing to the energy obtained by FCI (figure \ref{fig:H2PES}), all methods fit the curve pretty well. There are still some slight variations due to the number of terms in the ADAPT algorithms which is always lower than a full CI or even a UCC.

When comparing the error for Fermionic ADAPT and UCCSD (figure \ref{fig:H2PES}, one would expect the UCCSD error to always be lower than the Fermionic-ADAPT one. In fact, this is not always the case due to conditions over the convergence. Here are plot the best value depending on the parameters given to the optimizer and less parameters are given for the fermionic-ADAPT. We can expect UCCSD to have better result with more flexible conditions. The main issue stays the computation time.

On figure \ref{fig:H2ER}, one can see that Qubit and QEB ADAPT performs better than the Fermionic ADAPT and the UCCSD. Indeed, due to the larger pool of operators available, Qubit and QEB algorithms have more options to approximate the correct energy value.

\subsubsection{Number of CNOT and of parameters}

Because qiskit has many classes of object that are not always compatible, we coded QEB-Adapt with the \texttt{HamiltonianGate} \cite{hamham} class which needs a lot of computation time during the VQE, a better way would be to use \texttt{PauliEvolitionGate}\cite{hamiltoniangate_qiskit} used in Qubit-Adapt, wich is way faster to use in the VQE and allows to decompose the circuit, but the pauli string in the QEB-Adapt include complex coefficient wich are not allowed in the \texttt{PauliEvolutionGate}. A solution would have been to re-developped QEB-Adapt tailoring the \texttt{PauliEvolutionGate} framework. The \texttt{HamiltonianGate} does not allow us to decompose the circuit, thus we were not able to retrieve data on the number of parameters or CNOT gates as we did for the other methods.

As we can see in \ref{fig:H2PARAM}, all the methods have a constant number of parameters (number of operators added to the final ansatz) with the one of the Qubit-ADAPT being lower than the one of the UCCSD and the Fermionic-ADAPT. As expected the number of CNOT gates in the ansatz is lower for the Qubit-ADAPt than for the two other methods. It is also the only method that present variations in the number of CNOT gates and it seems interesting to remark that the drops (see \ref{fig:H2CNOT}) in number of CNOT gates are directly linked with the increase in the error that can be seen in the plot above.

\subsection{Results for $LiH$}

\subsubsection{Accuracy}

Concerning the QEB-ADAPT, we were not able to run it for any other molecule than $H_2$ because the number of operators in the pool grows too fast (for instance, it is around $1500$ for $LiH$) which made it way too long for us to compute (in \cite{yordanov_qubit-excitation-based_2021} they parallelize this computation).

As for $H_2$, one can see that each algorithm performs well with an error within the chemical accuracy. Due to very high computation time of these methods (around 24h), we had to put a threshold on the number of parameters (see figure \ref{fig:LIHPARAM}), thus we were not able to evaluate the full potential of Qubit-ADAPT and Ferm-ADAPT. Hence we see in \ref{fig:LIHER} that UCCSD with still all the parameters has a lower error.

\subsubsection{Number of CNOT and of parameters}

As explained above the number of operators in both ADAPT methods is thresholded due to computation time and is thus way smaller than the one for UCCSD. However it is interesting to remark that the Qubit-ADAPT has a lower error than the Fermionic one with a smaller number of CNOT gates.

\chapter*{Conclusion}

Through this project, we were able to tackle new fields such as quantum chemistry and the concept of Variational Quantum Algorithms, allowing us to discover possible near term applications of NISQ devices and so maybe providing feasible Quantum Advantage. While bringing a theoretical background on quantum chemistry, it allows to have enough knowledge to understand and approach the concepts of VQE, and in our case, especially ansatzes. Through the use of IBM's Qiskit we were able to develop and evaluate every ansatzes that we planned to implement from the simplest Hardware efficient to the hardest one, the QEB ADAPT ansatz. In the end we learned that, through efficient machine learning methods, the Qubit-ADAPT-VQE ansatz was able to bring high accuarcy while keeping a low number of parameters and reasonable number of CNOT gates (2-qubit gates being hard to implement),thus possibly being the most feasible ansatz.


\printbibliography

\end{document}